\documentclass[9pt,conference]{IEEEtran}
\IEEEoverridecommandlockouts

\usepackage{cite}
\usepackage{amsmath,amssymb,amsfonts}
\usepackage{algorithmic}
\usepackage{graphicx}
\usepackage{textcomp}
\usepackage{xcolor}
\usepackage{enumitem} 
\usepackage{multirow}
\usepackage{booktabs}
\usepackage{array} 
\usepackage{tabularx}
\usepackage{hyperref}
\usepackage{colortbl}
\usepackage{balance}
\usepackage{url}

\newcolumntype{L}[1]{>{\raggedright\let\newline\\\arraybackslash\hspace{0pt}}m{#1}}
\newcolumntype{R}[1]{>{\raggedleft\let\newline\\\arraybackslash\hspace{0pt}}m{#1}}
\newcolumntype{C}[1]{>{\centering\let\newline\\\arraybackslash\hspace{0pt}}m{#1}}

\def\BibTeX{{\rm B\kern-.05em{\sc i\kern-.025em b}\kern-.08em
    T\kern-.1667em\lower.7ex\hbox{E}\kern-.125emX}}

\begin{document}

\title{FLowHigh: Towards Efficient and High-Quality Audio Super-Resolution with Single-Step Flow Matching\\

\thanks{$^\dagger$Corresponding author.
This work was partly supported by the Institute of Information \& Communications Technology Planning \& Evaluation (IITP) grant funded by the Korea government (MSIT) (No. RS-2019-II190079, Artificial Intelligence Graduate School Program (Korea University), No. RS-2021-II-212068, Artificial Intelligence Innovation Hub, No. RS-2024-00336673, AI Technology for Interactive Communication of Language Impaired Individuals, and No. RS-2024-00436857, Information Technology Research Center (ITRC) support program).}
}

\author{\IEEEauthorblockN{1\textsuperscript{st} Jun-Hak Yun}
\IEEEauthorblockA{\textit{Department of Artificial Intelligence} \\
\textit{Korea University}\\
Seoul, Korea \\
jh\_yun@korea.ac.kr}
\and
\IEEEauthorblockN{2\textsuperscript{nd} Seung-Bin Kim}
\IEEEauthorblockA{\textit{Department of Artificial Intelligence} \\
\textit{Korea University}\\
Seoul, Korea \\
sb-kim@korea.ac.kr}
\and
\IEEEauthorblockN{3\textsuperscript{rd} Seong-Whan Lee$^\dagger$}
\IEEEauthorblockA{\textit{Department of Artificial Intelligence} \\
\textit{Korea University}\\
Seoul, Korea \\
sw.lee@korea.ac.kr}
}

\maketitle

\begin{abstract}
Audio super-resolution is challenging owing to its ill-posed nature. Recently, the application of diffusion models in audio super-resolution has shown promising results in alleviating this challenge. However, diffusion-based models have limitations, primarily the necessity for numerous sampling steps, which causes significantly increased latency when synthesizing high-quality audio samples. In this paper, we propose FLowHigh, a novel approach that integrates flow matching, a highly efficient generative model, into audio super-resolution. We also explore probability paths specially tailored for audio super-resolution, which effectively capture high-resolution audio distributions, thereby enhancing reconstruction quality. The proposed method generates high-fidelity, high-resolution audio through a single-step sampling process across various input sampling rates. The experimental results on the VCTK benchmark dataset demonstrate that FLowHigh achieves state-of-the-art performance in audio super-resolution, as evaluated by log-spectral distance and ViSQOL while maintaining computational efficiency with only a single-step sampling process. 
\end{abstract}


\begin{figure*}[htb!]
    \centering
        \includegraphics[width=\textwidth]{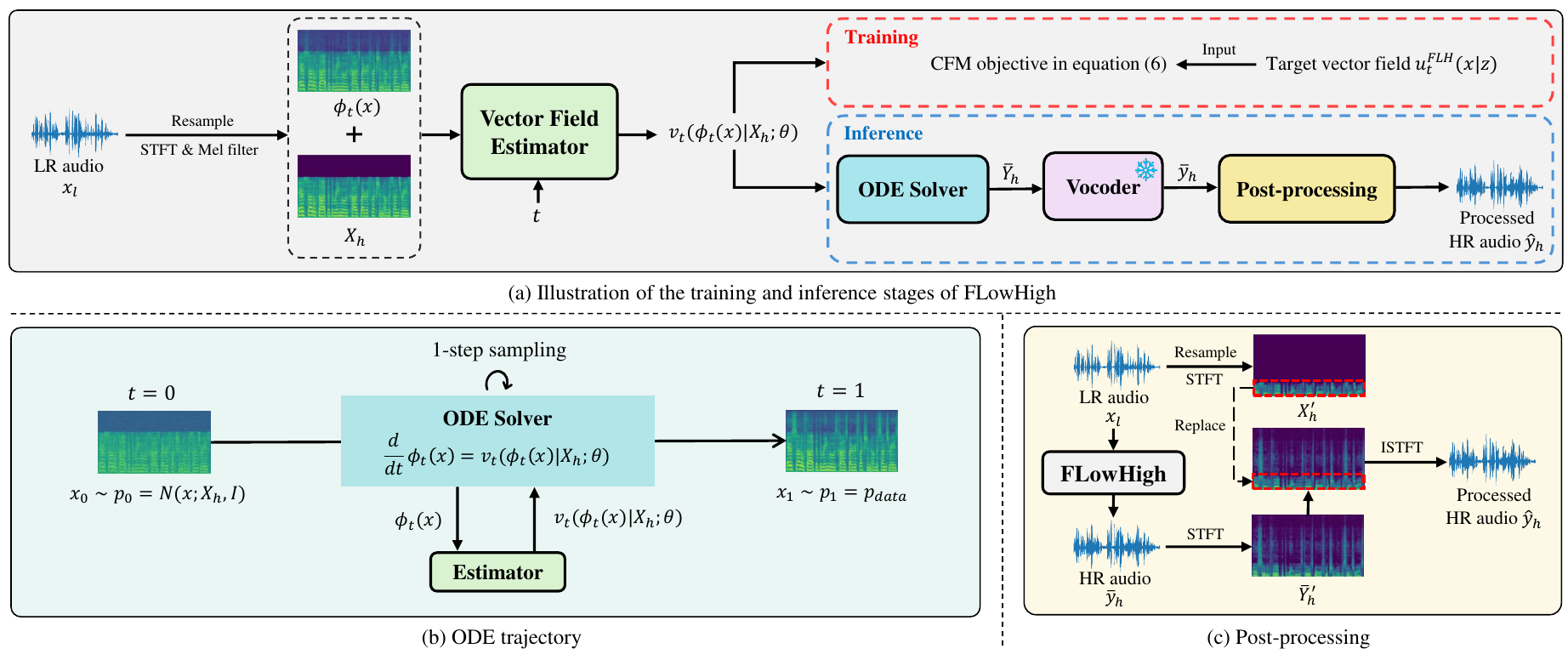}
        \vspace{-0.6cm}
        \caption{Overview of FLowHigh. (a) The overall training and inference process of FLowHigh based on conditional flow matching. (b) The ordinary differential equation trajectory begins from a data-dependent prior distribution. (c) Post-processing using STFT and ISTFT to replace the lower frequency components.}
        \vspace{-0.2cm}
        \label{fig:overview}
\end{figure*}

\section{Introduction}
Audio super-resolution (SR), also known as bandwidth extension, is a task with the purpose of reconstructing a high-resolution (HR) audio signal from a low-resolution (LR) input audio by estimating the missing high-frequency information, thereby enhancing perceptual quality.
Deep learning-based audio SR research \cite{ling2018waveform,abel2018simple,birnbaum2019temporal,wang2021towards} has greatly outperformed traditional approaches, building on advancements in deep learning and signal processing \cite{lee1996multiresolution,jeong2019classification,mane2020multi}.
Nevertheless, audio SR remains a challenging task due to its ill-posed nature, as an LR signal can correspond to numerous HR signals.

Recent works \cite{zhang2021wsrglow,lee2021nu,liu22y_interspeech} have employed generative models to address the one-to-many problem.
Several studies have leveraged generative adversarial networks (GANs), such as AERO \cite{10095382}, mdctGAN \cite{shuai23_interspeech}, and MS-BWE \cite{lu2024_interspeech}, which focus on spectrum estimation for inverse transformation.
Fre-painter \cite{kim2024audio} has demonstrated state-of-the-art performance by integrating a masked autoencoder and GANs with effective masking strategies.
While these studies have yielded promising results, GANs are prone to training instability and mode collapse, which can impede model convergence.

Recently, diffusion models \cite{NEURIPS2020_4c5bcfec, songdenoising,song2021scorebased} have demonstrated impressive performance in audio SR \cite{han2022nu,10095103,liu2024audiosr} as well as various audio generation tasks \cite{popov2021grad,liu2023audioldm, liu2024audioldm, byun2024midi,oh2024diffprosody}.
Nu-wave2 \cite{han2022nu} explores flexible input sampling rates and UDM+ \cite{10095103} introduces a conditional sampling algorithm to enhance low-frequency fidelity during the reverse process.
AudioSR \cite{liu2024audiosr} handles diverse audio types using a latent diffusion model. 
They also verify that leveraging prior knowledge from pre-trained vocoders can improve the audio SR performance \cite{liu22x_interspeech, liu2024audiosr}.
However, previous diffusion-based studies are often constrained by the need for a larger number of function evaluations (NFEs), leading to prolonged processing times to achieve high perceptual quality.
In real-world scenarios \cite{lee2019towards,won2020adaptive}, such limitations significantly impair efficiency and increase the latency for delay-sensitive operations.

Flow matching \cite{lipman2023flow, liu2023flow} is a simulation-free method for training continuous normalizing flows (CNFs) \cite{chen2018neural} with a simple vector field regression objective.
It learns the transformation between a simple prior and a complex data distribution via a straight trajectory, allowing for faster generation with fewer sampling iterations.
Recently, it has emerged as a promising approach, demonstrating exceptional performance and efficiency across diverse audio generation tasks \cite{le2023voicebox, vyas2023audiobox, liu2024generative, mehta2024matcha, prajwal2024musicflow, lee2024periodwave,cho2024emosphere++, jung2024flowavse}.
However, it has not yet been explored for audio SR.

Building on these insights, we propose FLowHigh, a novel audio SR method based on \textbf{F}low matching, developed to transform \textbf{Low}-resolution audio into \textbf{High}-resolution audio.
This study aims to address the shortcomings of diffusion-based models, concurrently elevating audio SR efficacy. 
By leveraging flow matching, FLowHigh models the HR audio distribution conditioned on LR input using a simple vector field regression objective without adversarial training.
FLowHigh incorporates a transformer-based vector field estimator that regresses the target vector field at the mel-spectrogram level. 
Additionally, it utilizes a pre-trained neural vocoder for synthesizing the waveform \cite{liu22x_interspeech}.
We further analyze the probability paths required for conditional flow matching (CFM) to devise a more efficient flow matching-based audio SR framework.
We found that well-defined path derived from input data are tailored for HR audio reconstruction.
Our approach notably outperforms existing methods and reconstructs the high-fidelity HR audio using a single-step sampling with the Euler method.
Consequently, it offers a substantial speed advantage over the computationally expensive sampling processes of diffusion models.

Our contributions are summarized as follows:
\begin{itemize}
    \item We propose FLowHigh, an efficient audio SR method based on flow matching. To the best of our knowledge, this is the first study to integrate flow matching into audio SR successfully.
    \item We analyze various probability paths essential for CFM, exploring different source distributions along the path, to refine and improve audio SR performance. 
    \item The experimental results demonstrate that FLowHigh outperforms existing models on objective metrics, achieving superior performance with only a single-step sampling across various audio sampling rates.
    Code implementations and audio samples are provided at 
    \url{https://jjunak-yun.github.io/FLowHigh}.
\end{itemize}

\section{Background: Flow Matching}
\label{sec:fm}
Let $x\in \mathbb{R}^d$ represent a data point in the data space $\mathbb{R}^d$, sampled from unknown distribution $q(x)$.
CNFs \cite{chen2018neural} are learned to transform a simple prior $p_0$ to a $p_1 \approx q$, where $q$ is a complex target distribution. 
CNFs define the time-dependent probability density function, called probability density path $p_{t} : [0,1] \times \mathbb{R}^{d} \to \mathbb{R}_{>0}$, where $\int p_t(x)dx =1$ and $t \in [0,1]$. 
The flow, time-dependent diffeomorphic mapping $\phi_{t}: [0,1] \times \mathbb{R}^d \to \mathbb{R}^d$, which induces the path $p_t$, is driven by time-dependent vector field $v_{t}: [0,1] \times \mathbb{R}^d \to \mathbb{R}^d$, defined via the following ordinary differential equation (ODE):
\begin{equation}
\frac{d}{dt} \phi_t(x)= v_t (\phi_t(x)), \quad \phi_0(x)=x.
\end{equation}

Flow matching \cite{lipman2023flow} has been proposed as a promising approach for simulation-free training CNFs.
The training objective of flow matching is simple vector field regression defined as follows:
\begin{equation} \label{eq:fmobjective}
\mathcal{L}_{FM} (\theta) = \mathbb{E}_{t\sim U[0,1],p_t(x)} ||u_t(x)-v_t(x;\theta)||^2_2,
\end{equation}
where $v_t(x;\theta)$ is the vector field parameterized by $\theta$  and $u_t(x)$ is a target vector field that produces the corresponding path $p_t$.
Since \eqref{eq:fmobjective} is intractable owing to the ignorance of the prior knowledge of $(u_t, p_t)$,
\cite{lipman2023flow} introduced tractable objective for CFM, and \cite{tong2024improving} further generalized it as:
\begin{equation} \label{eq:cfmobjective}
\mathcal{L}_{CFM} (\theta) = \mathbb{E}_{t\sim U[0,1],q(z),p_t(x|z)} || u_t(x|z) - v_t(x;\theta)||^2_2,
\end{equation}
where $p_t$ and $u_t$ are conditioned on $z$, which is sampled from some distribution $q$. 
They also proved that  $\mathcal{L}_{FM} (\theta)$ and $\mathcal{L}_{CFM} (\theta)$ have identical gradients for $\theta$ in the context of model optimization.
Assuming the probability path to be a time-dependent Gaussian conditional path $p_t(x|z)= \mathcal{N}(x|\mu_t(z), \sigma_t (z)^2 I)$, 
the target vector field that generates the flow $\phi_t(x) = \sigma_t(z)x + \mu_t(z)$ conditioned on $z$ is as follows: 
\begin{equation} \label{eq:targetVF}
u_t (x|z) = \frac{\sigma_t'(z)}{\sigma_t(z)} (x-\mu_t(z))+\mu_t' (z),
\end{equation}
where $\sigma_t'$ and $\mu_t'$ denote the derivatives of $\sigma_t$ and $\mu_t$, respectively.


\begin{table*}[!t]
    \centering
    \caption{Audio Super-Resolution Evaluation Results for 48 kHz with Input Sampling Rates of 8 kHz, 12 kHz, 16 kHz, and 24 kHz. The Mark $^{\mathrm{*}}$ Indicates Baselines with a Target Sampling Rate of 44.1 kHz. Numbers in \textbf{Bold} Represent the Best Results, While Numbers with an \underline{Underline} Represent the Sub-Optimal Results.} 
    \vspace{-0.2cm}
    \label{table1}
    \resizebox{\textwidth}{!}{%
\begin{tabular}{
@{} l @{} 
C{3.1em}@{\hskip 0.9\tabcolsep}C{4.0em}@{\hskip 0.7\tabcolsep}C{4.1em}@{\hskip 0.7\tabcolsep}C{4.1em} c @{} 
C{3.1em}@{\hskip 0.9\tabcolsep}C{4.0em}@{\hskip 0.7\tabcolsep}C{4.1em}@{\hskip 0.7\tabcolsep}C{4.1em} c @{} 
C{3.1em}@{\hskip 0.9\tabcolsep}C{4.0em}@{\hskip 0.7\tabcolsep}C{4.1em}@{\hskip 0.7\tabcolsep}C{4.1em} c @{} 
C{3.1em}@{\hskip 0.9\tabcolsep}C{4.0em}@{\hskip 0.7\tabcolsep}C{4.1em}@{\hskip 0.7\tabcolsep}C{4.1em} c @{}}

\toprule[0.7pt] 

& \multicolumn{4}{c}{\textbf{8 kHz $\rightarrow$ 48 kHz}} &
& \multicolumn{4}{c}{\textbf{12 kHz $\rightarrow$ 48 kHz}} &
& \multicolumn{4}{c}{\textbf{16 kHz $\rightarrow$ 48 kHz}} &
& \multicolumn{4}{c}{\textbf{24 kHz $\rightarrow$ 48 kHz}} \\

\cmidrule(lr){2-5} \cmidrule(lr){7-10} \cmidrule(lr){12-15} \cmidrule(l){17-20}

Model 

& \; {LSD$\downarrow$} & {LSD-LF$\downarrow$} & {LSD-HF$\downarrow$} & {ViSQOL$\uparrow$} &
& \; {LSD$\downarrow$} & {LSD-LF$\downarrow$} & {LSD-HF$\downarrow$} & {ViSQOL$\uparrow$} &
& \; {LSD$\downarrow$} & {LSD-LF$\downarrow$} & {LSD-HF$\downarrow$} & {ViSQOL$\uparrow$} &
& \; {LSD$\downarrow$} & {LSD-LF$\downarrow$} & {LSD-HF$\downarrow$} & {ViSQOL$\uparrow$} \\

\cmidrule(r){1-1} \cmidrule(lr){2-5} \cmidrule(lr){7-10} \cmidrule(lr){12-15} \cmidrule(l){17-20}

GT reconstruction 
& \; 0.71 & 0.66 & 0.76 & 4.67 &
& \; 0.71 & 0.66 & 0.76 & 4.67 &
& \; 0.71 & 0.66 & 0.76 & 4.67 &
& \; 0.71 & 0.66 & 0.76 & 4.67 \\

\quad w/ Post-processing
& \; 0.68 & 0.19 & 0.74 & 4.69 &
& \; 0.66 & 0.23 & 0.74 & 4.69 &
& \; 0.63 & 0.24 & 0.75 & 4.69 &
& \; 0.58 & 0.25 & 0.76 & 4.69  \\ 

\cmidrule(r){1-1} \cmidrule(lr){2-5} \cmidrule(lr){7-10} \cmidrule(lr){12-15} \cmidrule(l){17-20}

NVSR$^{\mathrm{*}}$ \cite{liu22x_interspeech}
& \; 0.99 & 0.49 & 1.07 & 3.05 & 
& \; 0.95 & 0.49 & 1.07 & 3.12 & 
& \; 0.90 & 0.47 & 1.07 & 3.26 &
& \; 0.83 & 0.47 & 1.11 & 3.66 \\

AudioSR$^{\mathrm{*}}$ \cite{liu2024audiosr}
& \; 1.35 & 0.39 & 1.47 & 2.31 &
& \; 1.24 & 0.47 & 1.42 & 2.36 &
& \; 1.18 & 0.48 & 1.43 & 2.45 &
& \; 0.94 & 0.49 & 1.27 & 2.92 \\

\cmidrule(r){1-1} \cmidrule(lr){2-5} \cmidrule(lr){7-10} \cmidrule(lr){12-15} \cmidrule(l){17-20}

Nu-wave2 \cite{han2022nu}

& \; 1.10 & 0.35 & 1.19 & 2.40 &   
& \; 0.95 & 0.36 & 1.07 & 2.70 &
& \; 0.86 & 0.37 & 1.02 & 3.00 &    
& \; 0.73 & 0.38 & 0.98 & 3.54 \\

UDM+ \cite{10095103}
& \; 1.27 & \textbf{0.11} & 1.35 & 2.29 &
& \; 1.08 & \textbf{0.14} & 1.23 & 2.51 &
& \; 0.94 & \textbf{0.14} & 1.15 & 2.77 &
& \; 0.75 & \textbf{0.14} & 1.05 & 3.48 \\

mdctGAN \cite{shuai23_interspeech}

& \; 1.02 & 0.27 & 1.10 & 2.69 &
& \; 0.98 & 0.31 & 1.11 & 2.59 &
& \; 0.87 & 0.29 & 1.04 & 2.83 &
& \; 0.75 & 0.30 & 1.06 & 3.48 \\

Fre-painter \cite{kim2024audio}

& \; 0.88 & 0.34 & 0.95 & 3.31 &
& \; 0.83 & 0.39 & 0.93 & 3.35 &
& \; \underline{0.77} & 0.37 & 0.90 & 3.54 &
& \; \underline{0.68} & 0.37 & 0.88 & 3.98 \\ 

\cmidrule(r){1-1} \cmidrule(lr){2-5} \cmidrule(lr){7-10} \cmidrule(lr){12-15} \cmidrule(l){17-20}

\textbf{FLowHigh}

& \; \textbf{0.81} & \underline{0.21} & \textbf{0.88} & \textbf{3.53} & 
& \; \textbf{0.75} & \underline{0.24} & \textbf{0.86} & \textbf{3.61} &
& \; \textbf{0.71} & \underline{0.25} & \textbf{0.85} & \textbf{3.80} &
& \; \textbf{0.62} & \underline{0.26} & \textbf{0.83} & \textbf{4.27} \\

\quad w/o Post-processing \;
& \; \underline{0.83} & 0.58 & \textbf{0.88} & \underline{3.50} & 
& \; \underline{0.80} & 0.62 & \textbf{0.86} & \textbf{3.61} &
& \; 0.78 & 0.64 & \textbf{0.85} & \underline{3.78} &
& \; 0.75 & 0.67 & \textbf{0.83} & \underline{4.24} \\

\toprule[0.7pt] 
\end{tabular}%
} \vspace{-0.3cm}
\end{table*}

\section{Methods}
\label{sec:methods}

We propose FLowHigh, a novel audio SR system utilizing flow matching at the mel-spectrogram level, as illustrated in Fig. \ref{fig:overview}. 
FLowHigh leverages the capacity of flow matching to competently model the complex distribution of HR audio representation data, enabling time-efficient and high-quality audio SR.  
More details are presented in the following subsections.

\subsection{Temporal and Spectral Processing of Input Signal} 
\label{ssec:pre-processing}

Following \cite{liu22x_interspeech}, given input LR signal $x_l=[x_{s_1}, x_{s_2}, ..., x_{S\cdot l}]$ with a sampling rate of $l$, we adjust the temporal resolution of the signal to generate $x_h=[x_{s_1}, x_{s_2}, ..., x_{S\cdot h}]$, where $S$ is the length of the audio signal in seconds and $h$ is the target sampling rate with $h>l$.  
To operate in the frequency domain, the complex spectrogram $X_h' \in \mathbb{R}^{N \times (\frac{n_{fft}}{2} + 1)} $ is computed from $x_h$ using a short-time Fourier transform (STFT) with $N$ frames, where $n_{fft}$ denotes the FFT size.
$X_h'$ is converted into a mel-spectrogram $X_h \in \mathbb{R}^{N \times F}$ by a mel-filter with $F$ bins, which serves as the input to our model. 
We denote that $y_h=[y_{s_1}, y_{s_2}, ..., y_{S\cdot h}]$ is the target HR signal.
Based on the Nyquist theorem, the highest bandwidths of the audio signals $x_l$, $x_h$, and $y_h$ are $l/2$, $l/2$, and $h/2$ Hz, respectively. 
Thus, we aim to reconstruct the missing high-frequency components of $x_h$ in the range $(l/2, h/2]$ Hz, approximating the target spectral representation of $y_h$. 

\subsection{Conditional Flow Matching for Audio Super-Resolution}
\label{ssec:fm}

When employing flow matching in audio SR, it is important to determine the appropriate conditional probability path $p_t(x|z)$, as this selection is influenced by the mean $\mu_t(z)$ and standard deviation $\sigma_t(z)$.
In this subsection, we introduce a probability path specifically designed to enhance the effectiveness of audio SR.
Extending the flow matching \cite{lipman2023flow}, the studies in \cite{tong2024improving, pooladian2023multisample} explored the generalized flow matching, which maps from an arbitrary source distribution rather than a standard Gaussian distribution to mitigate the independence of noise and the target data.
We align the source distribution with the LR audio data distribution to enhance the convergence ability to learn the flow.
This is achieved by strategically utilizing the mel-spectrogram obtained from the LR input audio, which helps improve the model’s capacity to effectively map between distributions.

Following simplified CFM of \cite{tong2024improving}, FLowHigh samples a pair of random variables $(x_0, x_1)$ as condition $z$ from $q(z)=q(x_0)q(x_1)$, where $x_0$ and $x_1$ denote the source point and target point. 
We let the conditional path with $\mu_t(z) = tx_1 + (1-t)x_0$, which is a linear interpolation between $x_0$ and $x_1$ over time.
Considering the absence of high-frequency components of data in the prior, we set $\sigma_t(z)=1-(1-\sigma)t$, where $\sigma$ is a sufficiently small value. 
Hence, FLowHigh is designed to model the straight path that transitions from a normal distribution centered at $x_0$ to $p_1(x|z)= \mathcal{N}(x|x_1,\sigma^2I)$. 
Within our framework, the condition $(x_0, x_1)$ represents the mel-spectrogram samples of the input and the ground truth HR waveform, respectively.
According to \eqref{eq:targetVF}, we can obtain the conditional target vector field formula for the FLowHigh training objective as follows:

\begin{equation}
u_t^{FLH}(x|z) =\frac{(x_1-x_0) - (1-\sigma)(x-x_0)}{1-(1-\sigma)t}.
\end{equation}

\subsection{Vector Field Estimator and Training Objective}
\label{ssec::vfestimator}

Inspired by \cite{le2023voicebox}, we adopt a transformer architecture \cite{NIPS2017_3f5ee243} as a vector field estimator to parameterize conditional vector field $v_t(x|X_h;\theta)$ for audio SR.
As depicted in Fig. \ref{fig:overview} (a), the model takes the mel-spectrogram $X_h$ and flow sample $\phi_t(x) \in \mathbb{R}^{N \times F}$ as an additional input, which is placed at the probability path at the flow step $t$.
Before being fed into the transformer, these inputs are concatenated and projected via a linear layer, resulting in an input of size $\mathbb{R}^{N \times d}$, where $d$ denotes the input dimension of the transformer block.
The transformer, conditioned on the flow step $t$, outputs the hidden representation, which is then passed through a final linear projection to predict the vector field $v_t$, which has the same dimension as the flow.

During training, we first obtain a sample $\phi_t(x)$ with randomly sampled flow step $t\sim U[0,1]$. 
Conditioned on $X_{h}$ from LR input, FLowHigh learns the flow that maps the sample $\phi_t(x)$ into the distribution of the target mel-spectrogram from HR audio by minimizing the vector field regression objective as follows:
\begin{equation}
    \mathcal{L} (\theta) = \mathbb{E}_{t, q(z), p_t(x|z)} || u_t^{FLH}(\phi_t(x)|z) - v_t(\phi_t(x)|X_h;\theta)||^2_2.
\end{equation}

\subsection{Solving the Ordinary Differential Equation for Inference}
\label{ssec:ODEsolver}
Once the estimator completes training, the source point $x_0$ is first drawn from $p_0$ to sample the high-fidelity mel-spectrogram $\bar{Y}_h$ of HR audio, which is rich in higher frequency information based on the $x_l$.
Subsequently, we use the estimator and the Euler ODE solver to numerically compute $x_1=\bar{Y}_h$, which is sampled from the learned distribution $p_1$, as illustrated in Fig. \ref{fig:overview} (b). 
The Euler method is as follows:
\begin{equation} \label{}
x_{t+\tau} = x_t + \tau v(x_t|X_h;\theta),
\end{equation}
where $\tau$ is the step size. 
In this study, we utilize only single-step sampling to obtain the $\bar{Y}_h$.
A pre-trained neural vocoder processes the $\bar{Y}_h$ to synthesize the waveform $\bar{y}_h$.
To retain the original lower frequency information of $x_l$ as completely as possible, we apply post-processing \cite{kim2024audio, liu22x_interspeech}, replacing lower frequency components of the $\bar{y}_h$ with those from the $x_l$ using STFT and inverse STFT (ISTFT), and obtain final output $\hat{y}_h \approx y_h$ as depicted in Fig. \ref{fig:overview} (c).

\section{Experiments}
\label{sec:typestyle}

\subsection{Dataset}
\label{sec:dataset}

We use the VCTK dataset \cite{yamagishi2019cstr}, which contains 44 hours of speech at a sampling rate of 48 kHz uttered by 108 English speakers, for both training and evaluation.
Following \cite{kim2024audio}, we divide the 108 speakers into a training set of 100 speakers and an evaluation set of 8 speakers.
Training data is generated by applying a Chebyshev Type I low-pass filter with random orders and ripples to the target audio, followed by downsampling to a lower sample rate $l$ randomly selected from 4 kHz to 32 kHz. For evaluation, the order 8 Chebyshev Type I low-pass filter with a ripple of  0.05 dB is applied to the target data. 
The audio SR experiments are conducted at various input sampling rates of 8, 12, 16, and 24 kHz, which correspond to cut-off frequencies of 4, 6, 8, and 12 kHz, respectively, for a target sampling rate of 48 kHz.

\subsection{Implementation Details}
\label{sec:implement}

The vector field estimator consists of 2 layers of transformer blocks with a 16-head self-attention layer, 1024 embedding dimensions, and 4096 dimensions for feed-forward networks, leading to 35.4M parameters.
We set the window size of 2048, hop length of 480, and FFT size of 2048 for STFT.
The waveform is synthesized using BigVGAN \cite{lee2023bigvgan}, a pre-trained neural vocoder that operates at 48 kHz with 256 mel bins, based on the official implementation and trained for 850k steps.
FLowHigh is trained with a batch size of 128 for 400k training steps on a single NVIDIA RTX A6000 GPU using the Adam optimizer \cite{KingBa15} with $\beta_1 = 0.9$, $\beta_2 = 0.99$, and $\epsilon = 10^{-8}$ and set the initial learning rate at $3\times10^{-4}$.
For high-quality reconstruction, we set $\sigma$ to $10^{-4}$.

\subsection{Evaluation Metrics}
\label{sec:Metrics}

We employ the log-spectral distance (LSD) \cite{han2022nu, liu22x_interspeech}, where lower values indicate better performance. 
Following \cite{kim2024audio}, we further calculate the low-frequency LSD (LSD-LF) and high-frequency LSD (LSD-HF) to assess the reconstruction performance across different frequency ranges.
We also utilize the virtual speech quality objective listener (ViSQOL) \cite{chinen2020visqol} to evaluate the perceptual quality in speech objectively, with higher scores indicating better perceived quality. 
We assess computational efficiency using real-time factor (RTF), calculated as the ratio of processing time to input audio's duration.

\section{Results} 
\label{sec:Results}

\begin{table}[t!]
    \caption{Comparison of Time-Efficiency and Audio Super-Resolution Evaluation Results Across Various Sampling Steps.}
    \vspace{-0.2cm}
    \label{table2}
    \centering
    \resizebox{0.98\columnwidth}{!}{
        \begin{tabular}{
            @{} l @{} 
            C{5.0em} @{} c @{\hskip 0.1\tabcolsep} 
            C{5.0em} @{} c @{\hskip 0.1\tabcolsep}
            C{4.1em} @{\hskip 0.5\tabcolsep} C{5.0em} @{} c @{}
            C{4.1em} @{\hskip 0.5\tabcolsep} C{5.0em} @{} 
            }
        \toprule
        
        & & & & & \multicolumn{2}{c}{\textbf{16 kHz $\to$ 48 kHz}} & & \multicolumn{2}{c}{\textbf{24 kHz $\to$ 48 kHz}} \\
        
        \cmidrule(lr){6-7} \cmidrule(l){9-10}
        
        Model & NFEs$\downarrow$ & & RTF$\downarrow$ & &  LSD$\downarrow$ & ViSQOL$\uparrow$ & & LSD$\downarrow$ & ViSQOL$\uparrow$ \\ 
        
        \cmidrule(r){1-1} \cmidrule(lr){2-2} \cmidrule(lr){4-4} \cmidrule(lr){6-7} \cmidrule(l){9-10}
        
        & 1 & & 0.0937 & & 2.56 & 1.62 & & 2.56 & 1.89  \\
        & 10 & & 0.2981 & & 1.24 & 2.36 & & 0.98 & 3.10  \\
        Nu-wave2 \cite{han2022nu} & 25 & & 0.6460 & & 0.95 & 2.73 & & 0.77 & 3.67  \\
        & 50 & & 1.2337 & & 0.86 & 3.00 & & 0.73 & 3.54  \\
        & 100 & & 2.4041 & & 0.85 & 2.99 & & 0.75 & 3.58  \\
        
        \cmidrule(r){1-1} \cmidrule(lr){2-2} \cmidrule(lr){4-4} \cmidrule(lr){6-7} \cmidrule(l){9-10}
        
        & 1 &  & 0.1796 &  & 2.54 & 1.61 & & 2.21 & 2.21  \\
        & 10 &  & 0.5646 &  & 1.41 & 2.30 & & 1.09 & 3.01  \\
        UDM+ \cite{10095103} & 25 &  & 1.1847 & & 1.13 & 2.59 & & 0.88 & 3.26  \\
        & 50 &  & 2.2415 &  & 0.94 & 2.77 & & 0.75 & 3.48  \\
        & 100 &  & 4.3212 &  & 0.84 & 2.92 & & 0.69 & 3.71  \\
        
        \cmidrule(r){1-1} \cmidrule(lr){2-2} \cmidrule(lr){4-4} \cmidrule(lr){6-7} \cmidrule(l){9-10}
        
        \textbf{FLowHigh} & 1 &  & 0.1769 & & \textbf{0.71} & \textbf{3.80} & & \textbf{0.62} & 4.27 \\ 
        \quad \; w/ Midpoint \; & 2 &  & 0.2527 & & \textbf{0.71} & \textbf{3.80} & & \textbf{0.62} & \textbf{4.28} \\

        \toprule 
        \end{tabular}
    }
\end{table}

\begin{table}[t!]
    
    \centering
    \caption{Conditional Probability Paths Analysis Results for Audio Super-Resolution From 16 kHz to 48 kHz. The Proposed Path Is Highlighted in Gray.}\vspace{-0.2cm}
    \label{table3}
    \resizebox{0.98\columnwidth}{!}{%

\begin{tabular}{ r | r |  r  | r |  c |  c  }
\toprule

\multicolumn{1}{r|}{$q(z)$} &
\multicolumn{1}{r|}{$p_0$} &
\multicolumn{1}{r|}{$\mu_t(z)$} & 
\multicolumn{1}{r|}{$\sigma_t$} &
\multicolumn{1}{c|}{\; LSD$\downarrow$\; } &
\multicolumn{1}{c}{ViSQOL$\uparrow$} \\ 

\midrule

$q(x_1)$ & $N(x|0,I)$ & $tx_1$ & $1-(1-\sigma)t$ & 
\textbf{0.71}  & 3.78  \\ [2pt] 


$q(x_0)q(x_1)$ & $N(x|X_h,\sigma^2 I)$ & $tx_1 + (1-t)x_0$ & $\sigma$ & 
0.73  & 3.75  \\ [2pt] 

\rowcolor[gray]{.9} $q(x_0)q(x_1)$ & $ N(x|X_h,I)$ & $tx_1 + (1-t)x_0$ & $1-(1-\sigma)t$ & 
\textbf{0.71}  & \textbf{3.80} \\ [2pt] 

\toprule 
\end{tabular}%
} \vspace{-0.3cm}

\end{table} 

\subsection{Evaluation of Audio Super-Resolution}
\label{ssec:obj_eval}

To evaluate the audio SR performance of FLowHigh, we compared it against several baselines, including Nu-wave2\footnote{We used the Nu-wave2 checkpoint from UDM+'s official implementation.}\cite{han2022nu}, mdctGAN \cite{shuai23_interspeech}, UDM+ \cite{10095103}, and Fre-painter \cite{kim2024audio}, all of which have exhibited notable performance in audio SR. 
NVSR \cite{liu22x_interspeech} and AudioSR \cite{liu2024audiosr} were included as additional baselines. These two models employ a two-stage process, comprising mel-spectrogram generation and waveform synthesis via a vocoder, which is similar to our procedure. 
For all baseline models, we used their official implementations and pre-trained checkpoints\footnote{Note that the official checkpoints of NVSR and \textit{speech} mode of AudioSR were trained with a target sampling rate of 44.1 kHz. For evaluation, we assessed these models using target audio at 44.1 kHz.} to ensure a fair comparison.
GT reconstruction represents the synthesized output from the ground truth mel-spectrogram using the same vocoder as in our model.

As shown in Table \ref{table1}, FLowHigh consistently demonstrated superior performance over the baseline methods, achieving the lowest LSD values across all evaluated input sampling rates.
Note that the GT reconstruction with post-processing serves as the upper bound for our model. 
Given 24 kHz audio as input, the output of our model was close to this upper bound, with an LSD difference of only 0.04.
Remarkably, these results were obtained using only single-step sampling, highlighting the inference efficiency of our model.
While UDM+ \cite{10095103} exhibited better restoration performance in the low-frequency range, our model was highly effective at reconstructing high-frequency components, as evidenced by the lowest LSD-HF values.
The results of the ViSQOL evaluation also showed that our model successfully generated high perceptual-quality audio across various source sampling rates.
Additionally, without post-processing, our model attained competitive LSD values and sub-optimal ViSQOL scores, leveraging the power of flow matching.

\subsection{Comparison by the Number of Sampling Steps}
\label{ssec:samplingstep}

\begin{figure}[t!]
    \centering
        \includegraphics[width=0.95\columnwidth]{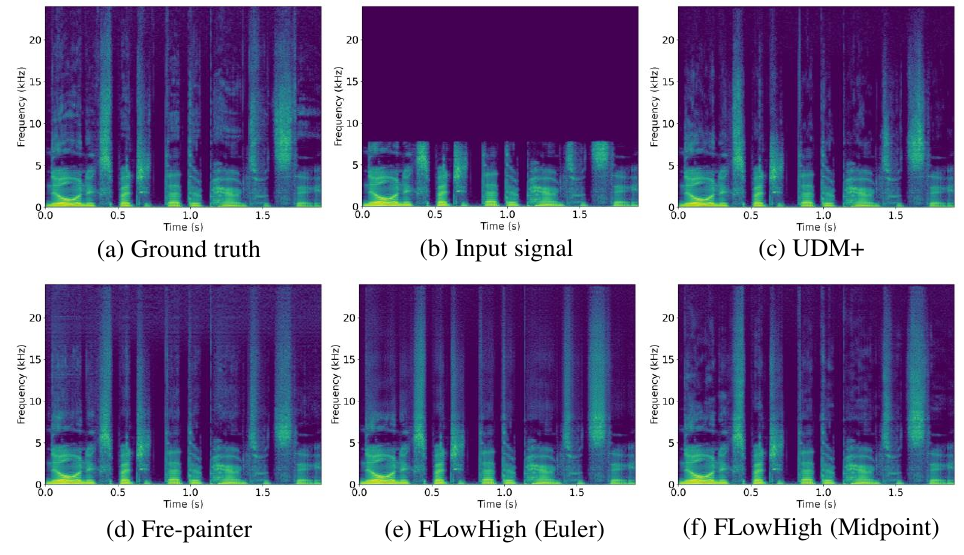}
        \vspace{-0.2cm}
        \caption{Spectrogram visualizations of ground truth, input signal, outputs from baselines, and FLowHigh for a target sample rate of 48 kHz. UDM+ uses 50 NFEs. The input sample rate of an audio sample (\textit{p360\_239}) is 16 kHz.}
        \vspace{-0.2cm}
        \label{fig:spectrogram}
\end{figure}

A significant drawback of diffusion models is that they require considerable NFEs to generate a high-quality sample.  
We conducted experiments using different numbers of sampling steps with diffusion-based models \cite{han2022nu, 10095103} to demonstrate the effectiveness and efficiency of FLowHigh, generating HR waveform at 48 kHz from input sampling rates of 16 and 24 kHz. 
Table \ref{table2} showed that diffusion-based models necessitate at least 50 NFEs to synthesize the high-quality samples. 
With only one NFE, our model achieved lower LSD and higher ViSQOL values than diffusion-based models running 100 steps, while also delivering inference speeds approximately 13.6 to 24.4 times faster.
While we successfully reconstructed audio using the Euler solver, we observed that adopting a midpoint method with 2 NFEs resulted in slightly higher fidelity, as shown in Fig. \ref{fig:spectrogram}.

\subsection{Analysis on Conditional Probability Path}
\label{ssec:path}

This analysis aims to advance the capabilities of a CFM-based audio SR framework by exploring suitable conditional probability paths.
We performed audio SR using various probability paths to validate the proposed path, with the evaluation results presented in Table \ref{table3}. 
First, we defined a path where $\mu_0(z)=0$ and $\mu_1(z) = x_1$, using the same $\sigma_t$ as the proposed path.
While starting from a standard Gaussian distribution yielded satisfactory reconstruction performance, we observed that a setting with a data-dependent prior resulted in comparatively higher perceptual quality.
We also conducted the experiments by defining the path with $\sigma_t(z) = \sigma$, where $\mu_t(z)$ is the same as in the proposed path.
This path had the advantage of allowing the target vector field $u_t$ to be expressed in the simplified formula $u_t(x|z) = x_1-x_0$. 
However, it resulted in a prior distribution with nearly empty high-frequency components, and this sparsity acted as an obstacle to learning the flow effectively in audio SR.

\section{Conclusion}
\label{sec:Conclusion}

In this work, we proposed FLowHigh, a novel audio SR model based on flow matching.
We leveraged the strength of CFM to effectively model the target data distribution for efficient audio reconstruction.
We also introduced suitable conditional paths tailored for audio SR, utilizing prior distributions corresponding to the input audio.
As a result, FLowHigh generated high-quality audio and outperformed existing audio SR methods across various input sampling rates in objective evaluations for a target sampling rate of 48 kHz.
Comparative experiments confirmed the inference efficiency of our model, as it produced high-fidelity audio with only single-step sampling, significantly accelerating the inference process compared to diffusion-based models.
However, there is room for improvement in audio quality through phase information modeling.
In future work, we plan to incorporate phase information modeling into advanced generative models to enhance the fidelity.

\vfill\pagebreak

\balance
\bibliographystyle{IEEEbib}
\bibliography{refs}

\end{document}